\documentclass[aps,preprint,showpacs,eqsecnum]{revtex4-1}
\usepackage{graphicx}
\usepackage{bm}
\usepackage{amsmath,upgreek}
\usepackage{hyperref}
\usepackage{txfonts}

\begin{document}

\title{Temperature effects on drift of suspended single-domain particles \\ induced by the Magnus force}
\author{S.~I.~Denisov}
\email{denisov@sumdu.edu.ua}
\author{T.~V.~Lyutyy}
\email{lyutyy@oeph.sumdu.edu.ua}
\author{V.~V.~Reva}
\author{A.~S.~Yermolenko}
\affiliation{Sumy State University, 2 Rimsky-Korsakov Street, UA-40007 Sumy, Ukraine}
%\date{submitted to Physical Review E: \today}

\begin{abstract}
We study the temperature dependence of the drift velocity of single-domain ferromagnetic particles induced by the Magnus force in a dilute suspension. A set of stochastic equations describing the translational and rotational dynamics of particles is derived, and the particle drift velocity that depends on components of the average particle magnetization is introduced. The Fokker-Planck equation for the probability density of magnetization orientations is solved analytically in the limit of strong thermal fluctuations for both the planar rotor and general models. Using these solutions, we calculate the drift velocity and show that the out-of-plane fluctuations of magnetization, which are not accounted for in the planar rotor model, play an important role. In the general case of arbitrary fluctuations, we investigate the temperature dependence of the drift velocity by numerically simulating a set of effective stochastic differential equations for the magnetization dynamics.
\end{abstract}
\maketitle

\section{INTRODUCTION}
\label{Intr}

Recently, the phenomenon of drift motion of single-domain ferromagnetic particles suspended in a viscous fluid has been analytically predicted and numerically confirmed \cite{DePe2017, DPKD2017}. This phenomenon arises from the Magnus effect, and drift of particles occurs if the external driving force, which induces their oscillatory motion, is properly synchronized with the rotating magnetic field, which induces their non-uniform rotation. The main characteristic of drift motion, the particle drift velocity, has been calculated both analytically and numerically. Because the magnitude and direction of this velocity are easy to operate, the drift phenomenon has been proposed to separate ferromagnetic particles in suspensions.

The approach used in \cite{DePe2017, DPKD2017} was purely deterministic, i.e., thermal fluctuations in the translational and rotational dynamics of particles were completely ignored. According to \cite{DePe2017}, this is approximately true if the particle size exceeds a few tens or even hundreds of nanometers. At the same time, to be single-domain, the particle size must be less than a critical one, which usually does not exceed hundreds of nanometers (see, e.g., Refs.~\cite{Skom2008, Gui2017}). As a consequence, the applicability of the deterministic approach is restricted only to relatively large single-domain particles. Therefore, in order to describe the drift motion of smaller particles, thermal fluctuations must be taken into account.

One of the most powerful tools to study the influence of thermal fluctuations on the behavior of dynamical systems is a set of stochastic differential equations (e.g., Langevin equations) in which the action of these fluctuations is modelled by Gaussian white noises \cite{HoLe1984, Gard2004, CKW2004}. The main advantage of these equations is that their solution is a Markov process, whose probability density function obeys the Fokker-Planck equation, see also Refs.~\cite{VanK2007, Risk1989}. Within this framework, the stochastic magnetization dynamics in fixed single-domain particles was first investigated by Brown \cite{Brow1963}. To describe this dynamics, he introduced the stochastic Landau-Lifshitz-Gilbert equation, in which thermal fluctuations are accounted for by a random magnetic field with Gaussian white noise components, and derived the corresponding Fokker-Planck equation. This approach (based on the stochastic Landau-Lifshitz-Gilbert or Landau-Lifshitz equations) has become a standard tool for studying the magnetization dynamics in nanosystems \cite{BMS2009}. In particular, it was employed to study the influence of uniformly rotating magnetic field on the nanoparticle magnetization \cite{DLH2006, DLHT2006}, magnetic relaxation \cite{DSTH2006, DSTH2007}, and precessional states of the magnetization \cite{DPL2011}.

A set of stochastic differential equations for the rotational motion of ferromagnetic particles in a viscous fluid depends on whether the magnetization vector is frozen into the particle body (see Ref.~\cite{CKW2004} and references therein) or not \cite{UsLi2012, UsUs2015, Usad2017, LHK2018}. In the former case, which holds if the anisotropy magnetic field noticeably exceeds the external one, the rotational dynamics of particles satisfies the Newton's second law for rotation with total torque containing Gaussian white noise contributions. This approach is widely used in studying the role of thermal fluctuations in the rotational dynamics of suspended particles \cite{CKW2004} (about the constructive role of these fluctuations see, e.g., Refs.~\cite{EMRJ2003, EnRe2004, LDRB2015}). But the role of temperature in the drift motion of single-domain ferromagnetic particles induced by the Magnus force has never been investigated before. Therefore, to fill this gap, in this work we develop a statistical theory of drift of ferromagnetic particles and determine the dependence of the particle drift velocity on temperature.

The outline of the paper is as follows. In Sec.~\ref{Set}, we derive a set of stochastic equations that describes the translational and rotational dynamics of single-domain ferromagnetic particles in a viscous fluid and accounts for the influence of thermal fluctuations. The particle drift velocity, which is induced by the Magnus force and depends on the average particle magnetization, is introduced in Sec.~\ref{Aver}. To find the average magnetization, in Sec.~\ref{FP-eq} we obtain the Fokker-Planck equation for the probability density function of magnetization orientations. In the same section, using this equation we propose a set of effective stochastic differential equations describing the magnetization dynamics, which is much simpler than the original one. In Sec.~\ref{Analyt}, in the case of strong thermal fluctuations we find analytically the steady-state solutions of the Fokker-Planck equations, which correspond to the planar rotor and general models, and on this basis calculate the drift velocity as a function of temperature for these models. In Sec.~\ref{Num}, we examine our theoretical predictions numerically and present numerical results on the temperature dependence of the drift velocity of $\mathrm{Co}$ nanoparticles in water, obtained by solving a set of effective stochastic equations. Finally, our main findings are summarized in Sec.~\ref{Concl}.

\section{Set of stochastic equations of motion}
\label{Set}

In our model, the ferromagnetic particles moving in a viscous fluid are considered to be spherical and smooth. Their radius $a$ is assumed to be so small that the single-domain state is realized, and thus the particle magnetization $\mathbf{M}$ is a function of time only: $\mathbf{M} = \mathbf{M} (t)$. If the anisotropy magnetic field is large enough, then the vector $\mathbf{M}$ is approximately parallel to the anisotropy axis. In this approximation the magnetization is frozen into the particle body, and hence its dynamics is governed by the kinematic differential equation $d\mathbf{M}/dt = \boldsymbol{ \omega} \times \mathbf{M}$, where $\boldsymbol{ \omega} = \boldsymbol{ \omega}(t)$ is the angular particle velocity. We also assume that the translational and rotational Reynolds numbers, $\mathrm{Re} _{t}$ and $\mathrm{Re}_{r}$, are small compared to 1 (since $\max{a}$ usually does not exceed a few hundreds of nanometers \cite{Skom2008, Gui2017}, this condition is not very restrictive for suspended particles). In this case, the inertial effects in the particle dynamics can be neglected and, as a consequence, the equations for the translational and rotational motions of particles are reduced to the force and torque balance equations, $\mathbf{F}=0$ and $\mathbf{T}=0$, respectively.

Neglecting interactions between particles, we include in the total force $\mathbf{F}$ acting on the particle the external driving force $\mathbf{f}_{d} = \mathbf{f}_{d}(t)$, the friction force $\mathbf{f}_{f} = \mathbf{f}_{f}(t)$, the Magnus lift force $\mathbf{f}_{l} = \mathbf{f}_{l}(t)$, and the random force $\mathbf{f} = \mathbf{f}(t)$. We choose the driving force in the form $\mathbf{f} _{d} = f_{m} \sin{(\Omega t - \phi)}\, \mathbf{e} _{x}$, where $f_{m}$, $\Omega$, and $\phi$ are, respectively, the amplitude, angular frequency, and initial phase of the force, and $\mathbf{e} _{x}$ is the unit vector along the $x$-axis of the Cartesian coordinate system. Under the above assumptions, the friction force is determined by the Stokes law, $\mathbf{f}_{f} = -6\pi\eta a\mathbf{v}$ ($\eta$ is the dynamic viscosity of the fluid, $\mathbf{v} = \mathbf{v} (t)$ is the linear particle velocity), and the Magnus force is given by \cite{RuKe1961} $\mathbf{f}_{l} = \pi\rho a^{3} \boldsymbol {\omega} \times \mathbf{v}$ ($\rho$ is the fluid density, the sign $\times$ denotes the vector product). Finally, we approximate the random force that accounts for thermal fluctuations by a Gaussian white noise with zero mean and correlation function (see, e.g., Ref.\ \cite{CKW2004})
\begin{equation}
    \langle f_{i}(t) f_{j}(t') \rangle =
    2\Delta_{1}\delta_{ij} \delta(t - t').
    \label{corr_f}
\end{equation}
Here, $f_{i}(t)$ ($i = x,y,z$) are the Cartesian components of the random force, the angular brackets denote averaging over all possible realizations of $\mathbf{f}(t)$, $\Delta_{1} = 6\pi \eta a k_{B} T$ is the noise intensity, $k_{B}$ is the Boltzmann constant, $T$ is the absolute temperature, $\delta_{ij}$ is the Kronecker symbol, and $\delta(\tau)$ is the Dirac $\delta$ function. Thus, in the considered case the force balance equation $\mathbf{f}_{d} + \mathbf{ f}_{f} + \mathbf{f}_{l} + \mathbf{f} = 0$ is stochastic and can be written in the form
\begin{equation}
    \mathbf{v} - \frac{\mathrm{Re}_{r}}
    {6\omega_{m}} \boldsymbol {\omega}
    \times \mathbf{v} = v_{m}\sin{
    (\Omega t - \phi)}\, \mathbf{e}_{x}
    + \frac{1} {6\pi \eta a}\mathbf{f},
    \label{v_eq}
\end{equation}
where the rotational Reynolds number is defined as $\mathrm{Re}_{r} = \rho a^{2}\omega_{m}/ \eta$, $\omega_{m} = \max{|\boldsymbol {\omega}|}$, and $v_{m} = f_{m}/6\pi \eta a$. (For completeness, the translational Reynolds number is defined as $\mathrm{Re}_{t} = \rho av_{m}/ \eta$.)

Next, we assume that the suspended particles are subjected to a homogeneous external magnetic field $\mathbf{H} = \mathbf{H}(t)$. Since the torque exerted on the particle magnetic moment $V \mathbf{M}$ ($V = 4\pi a^{3}/3$ is the particle volume) by this field equals $V \mathbf{M} \times \mathbf{H}$ and the magnetization is considered to be frozen, the mechanical torque $\mathbf{t}_{m}$ induced by the magnetic field is given by $\mathbf{t}_{m} = V\mathbf{M} \times \mathbf{H}$. In addition to this contribution, the total torque $\mathbf{T}$ acting on the particle includes also the frictional torque $\mathbf{t}_{f} = -8\pi \eta a^{3} \boldsymbol {\omega}$ and the random torque $\mathbf{t} = \mathbf{t}(t)$, which is approximated by another Gaussian white noise with $\langle t_{i}(t) \rangle = 0$ and
\begin{equation}
    \langle t_{i}(t) t_{j}(t') \rangle =
    2\Delta_{2}\delta_{ij} \delta(t - t'),
    \label{corr_t}
\end{equation}
where $\Delta_{2} = 6\eta Vk_{B}T$ is the intensity of this noise \cite{CKW2004}. Therefore, the torque balance equation takes the form $\mathbf{t}_{m} + \mathbf{t}_{f} + \mathbf{t}= 0$, and its solution with respect to the particle angular velocity yields
\begin{equation}
    \boldsymbol{\omega} = \frac{1}{6\eta}
    \mathbf{M} \times \mathbf{H} + \frac{1}
    {6\eta V}\mathbf{t}.
    \label{omega}
\end{equation}
Substituting this expression for the angular velocity into the kinematic equation, we obtain the following stochastic differential equation for the particle magnetization:
\begin{equation}
    \frac{d}{dt}\mathbf{M} = -\frac{1}{6\eta}
    \mathbf{M} \times (\mathbf{M} \times
    \mathbf{H}) - \frac{1}{6\eta V}\mathbf{M}
    \times \mathbf{t}.
    \label{M_eq}
\end{equation}

Thus, in our model, the translational and rotational motions of suspended ferromagnetic particles are described by the stochastic equations (\ref{v_eq}), (\ref{omega}), and (\ref{M_eq}). They show that, while the particle rotation influences (due to the Magnus force) the translational motion, the rotational motion does not depend on the translational one. Therefore, to find statistical characteristics of the translational motion of these particles, including their drift velocity, we must first determine the rotational properties that are described by Eqs.~(\ref{omega}) and (\ref{M_eq}).

\subsection{Set of stochastic equations for the magnetization dynamics}
\label{Magn}

As a first step, let us rewrite Eq.~(\ref{M_eq}) in the dimensionless form. Introducing the dimensionless time $\tau = \Omega t/2\pi$, magnetization $\mathbf{m} = \mathbf{m}(\tau) = \mathbf{M}(2\pi \tau/ \Omega)/M$, external magnetic field $\mathbf{h} = \mathbf{h}(\tau) = \mathbf{H}(2\pi \tau/ \Omega) /H_{m}$ ($H_{m} = \max{|\mathbf{H}|}$), and Gaussian white noise $\boldsymbol{\xi} = \boldsymbol{\xi}(\tau) = \mathbf{t}(2\pi \tau/\Omega)/ \sqrt{4\eta a^{3}\Omega k_{B}T}$, one obtains
\begin{equation}
    \dot{\mathbf{m}} = -\alpha \mathbf{m}
    \times (\mathbf{m} \times \mathbf{h})
    -\beta \mathbf{m}\times\boldsymbol{\xi}.
    \label{m_eq}
\end{equation}
Here, the overdot denotes the derivative with respect to $\tau$, the mean and correlation function of $\boldsymbol{\xi}$ are given by
\begin{equation}
    \langle \xi_{i}(\tau) \rangle =0, \qquad
    \langle \xi_{i}(\tau) \xi_{j}(\tau')
    \rangle= 2\delta_{ij}\delta(\tau- \tau'),
    \label{cor_xi}
\end{equation}
and the dimensionless parameters $\alpha$ and $\beta$ are as follows:
\begin{equation}
    \alpha = \frac{\pi MH_{m}}{3 \eta \Omega},
    \qquad
    \beta = \left( \frac{k_{B}T}{4\eta a^{3}
    \Omega} \right)^{1/2}.
    \label{def_a,b}
\end{equation}
According to Eq.~(\ref{m_eq}), the parameter $\alpha$ can be associated with the inverse rotational relaxation time, and the parameter $\beta$ characterizes the magnitude of thermal fluctuations. Note also that in these variables the particle angular velocity (\ref{omega}) reads
\begin{equation}
    \boldsymbol{\omega} = \frac{\Omega}{2\pi}
    (\alpha \mathbf{m} \times \mathbf{h} +
    \beta \boldsymbol{\xi}).
    \label{omega2}
\end{equation}

Next, we represent the unit vector $\mathbf{m}$ in the form
\begin{equation}
    \mathbf{m} = \sin{\theta} \cos{\varphi}
    \, \mathbf{e}_{x} + \sin{\theta}
    \sin{\varphi}\,\mathbf{e}_{y} + \cos{\theta}
    \,\mathbf{e}_{z},
    \label{def_m}
\end{equation}
where $\theta = \theta(\tau)$ and $\varphi = \varphi(\tau)$ are, respectively, the polar and azimuthal angles of $\mathbf{m}$. In addition, we assume that the external magnetic field has a fixed magnitude $|\mathbf{H}| = H_{m}$ ($|\mathbf{h}|=1$), lies in the $xy$ plane, and non-uniformly rotates around the $x$ axis, i.e.,
\begin{equation}
    \mathbf{h} = \cos{\psi}\,\mathbf{e}_{x}
    + \sin{\psi}\,\mathbf{e}_{y},
    \label{def_h}
\end{equation}
where $\psi = \psi(\tau)$ is the azimuthal angle of $\mathbf{h}$. Now, substituting (\ref{def_m}) and (\ref{def_h}) into Eq.~(\ref{m_eq}), we get the following set of stochastic equations for the polar and azimuthal angles of the magnetization vector:
\begin{equation}
\begin{split}
    \dot{\theta} &= \alpha \cos{\theta}
    \cos{(\psi - \varphi)} - \beta(\xi_{x}
    \sin{\varphi} - \xi_{y}\cos{\varphi}),
    \\
    \dot{\varphi} &= \alpha \frac{\sin{
    (\psi - \varphi)}}{\sin{\theta}} -
    \beta \cot{\theta} (\xi_{x}\cos{\varphi}
    + \xi_{y}\sin{\varphi}) + \beta \xi_{z}.
    \label{set}
\end{split}
\end{equation}
Here, the angle $\psi$ is a given function of the dimensionless time $\tau$, which, similar to the driving force, is assumed to satisfy the antisymmetry condition $\psi(\tau  + 1/2) = -\psi(\tau)$. It should also be stressed that this condition implies that $\psi(\tau)$ is a periodic function with period 1.

For convenience, let us introduce a new variable, the lag angle $\chi = \psi - \varphi$, instead of the azimuthal angle $\varphi$. Then Eqs.~(\ref{set}) can be rewritten in the form
\begin{equation}
\begin{split}
    \dot{\theta} &= \alpha \cos{\theta}
    \cos{\chi} - \beta[\xi_{x} \sin{(\psi
    - \chi)} - \xi_{y}\cos{(\psi - \chi)}],
    \\
    \dot{\chi} &= \dot{\psi} - \alpha
    \frac{\sin{\chi}}{\sin{\theta}} +\! \beta
    \cot{\theta} [\xi_{x}\cos{(\psi - \chi)}
    + \xi_{y}\sin{(\psi - \chi)}] - \beta \xi_{z}.
    \label{set2}
\end{split}
\end{equation}
It is these stochastic differential equations that we will use to study the effect of thermal fluctuations on the drift of particles.

\subsection{Random particle velocity}
\label{Vel}

In principle, Eq.~(\ref{v_eq}) can be solved exactly with respect to the dimensionless particle velocity $\mathbf{u} = \mathbf{u} (\tau) = \mathbf{v}(2\pi \tau/ \Omega) /v_{m}$. However, since the Magnus force is calculated with linear accuracy in Reynolds numbers \cite{RuKe1961}, the solution of this equation should be determined with the same accuracy. Therefore, representing $\mathbf{u}$ in the form $\mathbf{u} = \mathbf{u}_{0} + \mathbf{u}_{1}$, where $|\mathbf{u}_{0}| \sim 1$ and $|\mathbf{u} _{1}| \sim \mathrm{Re}_{r}$, from Eq.~(\ref{v_eq}) we immediately find
\begin{equation}
    \mathbf{u}_{0} = \sin{(2\pi\tau - \phi)}
    \,\mathbf{e}_x + \sigma \boldsymbol{\nu}
    \label{u_0}
\end{equation}
and
\begin{equation}
    \mathbf{u}_{1} = \frac{\mathrm{Re}_{r}}
    {6\omega_{m}} \boldsymbol{\omega} \times
    \mathbf{u}_{0}.
    \label{u_1}
\end{equation}
Here, $\sigma = \sqrt{\Omega k_{B}T /2\pi f_{m}v_{m}}$ is the dimensionless parameter characterizing the magnitude of the random force fluctuations, and $\boldsymbol{\nu} = \boldsymbol{\nu}(\tau) = \mathbf{f}(2\pi \tau/\Omega)/ \sqrt{3\eta a\Omega k_{B}T}$ is a dimensionless Gaussian white noise with
\begin{equation}
    \langle \nu_{i}(\tau) \rangle =0,\qquad
    \langle \nu_{i}(\tau) \nu_{j}(\tau')
    \rangle= 2\delta_{ij}\delta(\tau- \tau').
    \label{cor_nu}
\end{equation}
Thus, according to (\ref{u_0}) and (\ref{u_1}), the random particle velocity can be written as
\begin{align}
    \mathbf{u} = & \left(\mathbf{e}_x +
    \frac{\mathrm{Re}_{r}} {6\omega_{m}}
    \boldsymbol{\omega}\times \mathbf{e}_{x}
    \right) \sin{(2\pi\tau - \phi)}
    \nonumber \\
    & + \sigma \left(\boldsymbol{\nu} +
    \frac{\mathrm{Re}_{r}} {6\omega_{m}}
    \boldsymbol{\omega}\times \boldsymbol{\nu}
    \right).
    \label{u}
\end{align}

It is important to note that expression (\ref{u}) is valid only if the particle angular velocity $\boldsymbol{ \omega}$ does not contain a white-noise contribution. The reason is that the multiplication of generalized functions, including Gaussian white noises, is not mathematically defined (for more details see, e.g., Ref.~\cite{HoLe1984}). Therefore, if $\boldsymbol{ \omega}$ is given by (\ref{omega2}), then the particle velocity (\ref{u}) is not appropriate for studying the stochastic translational motion of suspended particles (since $\mathbf{u}$ contains the term proportional to $\boldsymbol{\xi} \times \boldsymbol{\nu}$). However, expression (\ref{u}) can be used to find the drift velocity of these particles, see just below.

\section{Particle drift velocity}
\label{Aver}

In order to derive the expression for the particle drift velocity, let us assume for the moment that the angular velocity is governed by the stochastic differential equation $J d\boldsymbol{ \omega}/dt = \mathbf{T}$ ($J$ is the moment of inertia of the particle). In this case, the above problem does not occur and the particle velocity (\ref{u}) can be averaged with respect to realizations of white noises $\boldsymbol{\xi}$ and $\boldsymbol{\nu}$. Taking into account that $\langle \boldsymbol{\nu} \rangle = 0$ and, since components of these noises are statistically independent, the condition $\langle \boldsymbol{ \omega} \times \boldsymbol{\nu} \rangle = \langle \boldsymbol{ \omega} \rangle \times \langle \boldsymbol{\nu} \rangle = 0$ holds, we obtain
\begin{equation}
    \langle \mathbf{u} \rangle = \left(
    \mathbf{e}_x + \frac{\mathrm{Re}_{r}}
    {6\omega_{m}}\langle \boldsymbol{
    \omega} \rangle \times \mathbf{e}_{x}
    \right) \sin{(2\pi\tau - \phi)}.
    \label{<u>1}
\end{equation}
Now, using the inertialess approximation ($J \to 0$) that leads to the torque balance equation $\mathbf{T} =0$, from (\ref{<u>1}) and (\ref{omega2}) one finds the following expression for the instantaneous mean particle velocity:
\begin{equation}
    \langle \mathbf{u} \rangle = \left[
    \mathbf{e}_x - \gamma\, \mathbf{e}_x
    \times (\langle\mathbf{m}\rangle
    \times \mathbf{h}) \right] \sin{(2\pi
    \tau - \phi)},
    \label{<u>2}
\end{equation}
where the dimensionless parameter
\begin{equation}
    \gamma = \frac{\mathrm{Re}_{r} \alpha
    \Omega}{12 \pi \omega_{m}} =
    \frac{\rho a^{2}MH_{m}}{36\eta^{2}}
    \label{def_g}
\end{equation}
($\gamma \ll 1$) characterizes the contribution of the Magnus force to $\langle \mathbf{u} \rangle$. Expression (\ref{<u>2}) shows that thermal fluctuations influence the mean particle velocity only through the dimensionless mean magnetization $\langle \mathbf{m} \rangle$ of the particle.

We define the dimensionless drift velocity $\langle \mathbf{s} \rangle$ of the particle as its steady-state displacement during one magnetic field period, i.e., $\langle \mathbf{s} \rangle = \lim_{n \to \infty} \int_{n}^{n+1} \langle \mathbf{u}(\tau) \rangle\, d\tau$ ($n$ is a whole number). In order to find an explicit expression for $\langle \mathbf{s} \rangle$, we first note that, according to (\ref{def_m}) and (\ref{def_h}),
\begin{equation}
    \mathbf{e}_x \times (\langle\mathbf{m}
    \rangle \times \mathbf{h}) = -
    \langle\sin{\theta} \sin{\chi}\rangle
    \,\mathbf{e}_{y} + \langle\cos{\theta}
    \rangle \cos{\psi}\,\mathbf{e}_{z}.
    \label{prod}
\end{equation}
Then, putting $\tau = n + \xi$ with $\xi \in (0,1)$, one can write $\lim_{n \to \infty} \langle \mathbf{u}(n + \xi) \rangle = \langle \mathbf{u}_{\mathrm{st}}(\xi) \rangle$, where the subscript `st' refers to the steady state. Finally, using the symmetry-caused condition $\langle \cos{\theta_{\mathrm{st}} (\xi)}\rangle = 0$ (for its proof see the next section) and the above results, one obtains $\langle \mathbf{s} \rangle = s_{y}\, \mathbf{e}_{y}$ and
\begin{equation}
    \langle s_{y} \rangle = \gamma
    \int_{0}^{1} \langle \sin{\theta_{
    \mathrm{st}}(\xi)} \sin{\chi_{\mathrm{st}}
    (\xi)}\rangle \sin{(2\pi \xi - \phi)}\,d\xi.
    \label{<s_y>1}
\end{equation}

Thus, the drift motion of particles induced by the Magnus force occurs along the $y$ axis with the drift velocity (\ref{<s_y>1}). In general, this velocity can be calculated by numerically solving a set of coupled stochastic equations (\ref{set2}) describing the magnetization dynamics. But to determine the drift velocity analytically, it is convenient to introduce the Fokker-Planck equation associated with these stochastic equations.

\section{Fokker-Planck equation for the magnetization dynamics}
\label{FP-eq}

The Fokker-Planck equation, which corresponds to a set of stochastic differential equations (\ref{set2}), can be obtained by standard techniques \cite{HoLe1984, Gard2004, CKW2004, VanK2007, Risk1989}. An important feature of these equations is that Gaussian white noises are multiplicative and, as a consequence, the statistical properties of their solution, $\theta (\tau)$ and $\varphi (\tau)$, depend on the interpretation of these noises. Using an arbitrary interpretation, in Ref.~\cite{LDRB2015} we have shown that noises $\xi_{i}(\tau)$ must be interpreted in the Stratonovich sense \cite{Stra1966} (to get Boltzmann distribution in equilibrium) and have derived the corresponding Fokker-Planck equation in the case of uniformly rotating magnetic field. A simple generalization of our approach to the case of non-uniformly rotating magnetic field leads to the Fokker-Planck equation which, in the operator form, can be written as
\begin{equation}
    \frac{\partial}{\partial\tau} P
    + \dot{\psi} \frac{\partial}{
    \partial\chiup} P = \hat{L} P.
    \label{FP}
\end{equation}
Here, $P = P(\thetaup, \chiup, \tau)$ is the probability density that $\theta (\tau) = \thetaup$ ($0 \leq \thetaup \leq \pi$) and $\chi (\tau) = \chiup$ ($0 \leq \chiup < 2\pi$), i.e.,
\begin{equation}
    P = \langle \delta{[\thetaup -
    \theta(\tau)]}\, \delta{[\chiup -
    \chi (\tau)]} \rangle,
    \label{def_P}
\end{equation}
and
\begin{align}
    \hat{L} P = &- \alpha \cos{\chiup}
    \frac{\partial} {\partial \thetaup}
    \cos{\thetaup} P + \alpha \frac{1}{
    \sin{\thetaup}}\frac{\partial}{
    \partial \chiup}\sin{\chiup} P
    \nonumber \\
    &  - \beta^{2}\frac{\partial}{\partial
    \thetaup} \cot{\thetaup} P + \beta^{2}
    \frac{\partial^{2}}{\partial \thetaup^{2}}
    P + \beta^{2}\frac{1} {\sin^{2}{\thetaup}}
    \frac{\partial^{2}}{\partial \chiup^{2}}P.
    \label{def_L}
\end{align}
As usually, $P$ should satisfy the normalization condition $\int_{0}^ {\pi} \int_{0}^{2\pi} P\, d\thetaup d\chiup =1$ and the initial condition $P(\thetaup, \chiup, 0) = \delta(\thetaup - \theta_{0})\, \delta(\chiup - \chi_{0})$, where $\theta_{0} = \theta(0)$ and $\chi_{0} = \chi(0)$.

Now, let us prove that $\langle \cos{\theta_{ \mathrm{st}} (\xi)}\rangle =0$. To this end, we first define the steady-state probability density as $P_{ \mathrm{st}}(\thetaup, \chiup, \xi) = \lim_{n \to \infty} P (\thetaup,\chiup, n + \xi)$ and represent $\langle \cos{\theta_{ \mathrm{st}} (\xi)}\rangle$ in the form
\begin{equation}
    \langle \cos{\theta_{\mathrm{st}}
    (\xi)}\rangle = \int_{0}^ {\pi}
    \int_{0}^{2\pi} \cos{\thetaup}
    P_{\mathrm{st}}(\thetaup,\chiup,
    \xi)\, d\thetaup d\chiup.
    \label{<cos>}
\end{equation}
Then, using the symmetry condition $P_{ \mathrm{st}} (\thetaup, \chiup, \xi) = P_{ \mathrm{st}} (\pi - \thetaup, \chiup, \xi)$, which follows from the Fokker-Planck equation (\ref{FP}), one can easily make sure that $\int_{0}^ {\pi} \cos{\thetaup} P_{ \mathrm{st}} (\thetaup, \chiup, \xi)\, d\thetaup = 0$ and, hence, the condition $\langle \cos{\theta_{ \mathrm{st}} (\xi)}\rangle =0$ indeed holds.

\subsection{Set of effective stochastic equations for the magnetization dynamics}
\label{Red_set}

Recall, the Fokker-Planck equation (\ref{FP}) corresponds to a set of stochastic differential equations (\ref{set2}), in which  Gaussian white noises $\xi_{i}(\tau)$ are interpreted in the Stratonovich sense. It is not difficult to verify (see, e.g., Ref.~\cite{ LDRB2015}) that the same Fokker-Planck equation corresponds also to the following set of effective stochastic differential equations:
\begin{equation}
\begin{split}
    \dot{\theta} &= \alpha \cos{\theta}
    \cos{\chi} + \beta^{2} \cot{\theta}
    + \sqrt{2}\, \beta \zeta_{1},
    \\
    \dot{\chi} &= \dot{\psi} - \alpha
    \frac{\sin{\chi}}{\sin{\theta}} -
    \sqrt{2}\, \beta \frac{1}{\sin{
    \theta}} \zeta_{2},
    \label{red_set}
\end{split}
\end{equation}
where Gaussian white noises $\zeta_{i} = \zeta_{i}(\tau)$, ($i = 1,2$) are interpreted in the It\^{o} sense \cite{Ito1950} and are characterized by
\begin{equation}
    \langle \zeta_{i}(\tau) \rangle = 0,
    \qquad \langle \zeta_{i}(\tau) \zeta_{j}
    (\tau') \rangle = \delta_{ij} \delta
    (\tau - \tau').
    \label{corr2}
\end{equation}
This means that solutions of Eqs.~(\ref{set2}) and (\ref{red_set}) with respect to the polar and lag angles $\theta(\tau)$ and $\chi(\tau)$ have the same probability density $P(\thetaup, \chiup, \tau)$. The word `effective' is used here to emphasize that a set of Eqs.~(\ref{red_set}) (i) is not a set of ordinary stochastic differential equations (because the noiseless term $\beta^{2} \cot{\theta}$ arises, nevertheless, from thermal fluctuations) and (ii) is much simpler than a set of basic equations (\ref{set2}).

An interesting property of Eqs.~(\ref{red_set}) that follows from the results of Ref.~\cite{MDCW2014} is that, in spite of the multiplicative character of the noise $\zeta_{2}$, the statistical characteristics of their solution do not depend on interpretation of noises $\zeta_{i}$ \cite{ LDRB2015}. Note also that sets of effective stochastic  differential equations describing the rotational dynamics of ferromagnetic particles in a viscous fluid and the magnetization dynamics in ferromagnetic particles embedded into a solid matrix have been proposed in Refs.~\cite{RaEn2004} and \cite{DSTH2007}, respectively.

\section{Analytical results: Strong thermal fluctuations}
\label{Analyt}

Because small thermal fluctuations do not affect strongly the deterministic drift of particles, which was studied in detail in Refs.~\cite{DePe2017, DPKD2017}, below we analytically study the role of large thermal fluctuations only. For this purpose, we consider the stochastic particle dynamics within the framework of the planar rotor model and general model.

\subsection{Planar rotor model}
\label{Rotor}

Within this model, the magnetization rotation is assumed to occur only in the $xy$ plane. This rotation is described by the second equation in (\ref{red_set}) with $\theta = \pi/2$, i.e.,
\begin{equation}
    \dot{\chi} = \dot{\psi} - \alpha
    \sin{\chi} -\sqrt{2}\,\beta \zeta_{2}.
    \label{eq_chi}
\end{equation}
The Fokker-Planck equation associated with Eq.~(\ref{eq_chi}) can be written as follows [cf.\ Eq.~(\ref{FP})]:
\begin{equation}
    \frac{\partial}{\partial \tau} P
    + \dot{\psi} \frac{\partial}{\partial
    \chiup} P = \hat{L}_{\bot} P,
    \label{FP3}
\end{equation}
where $P = P(\chiup, \tau)$ is the probability density that $\chi (\tau) = \chiup$ and
\begin{equation}
    \hat{L}_{\bot} P = \alpha \frac{
    \partial}{\partial \chiup} \sin{
    \chiup} P + \beta^{2} \frac{\partial
    ^{2}} {\partial\chiup^{2}} P.
    \label{L_perp}
\end{equation}

In equilibrium, when $\dot{\psi} = 0$ ($\psi(\tau) = \mathrm{ const}$) and $\tau \to \infty$, i.e., when the left-hand side of the Fokker-Planck equation (\ref{FP3}) equals zero, the equilibrium probability density $P_{\mathrm{eq}} (\chiup)$ satisfies the equation $\hat{L}_{\bot} P_{ \mathrm{eq}} (\chiup) = 0$. It can easily be verified that its normalized solution on any interval of length $2\pi$ [e.g., $\chi \in [0, 2\pi)$] is the von Mises probability density function
\begin{equation}
    P_{\mathrm{eq}}(\chiup) = \frac{e^{
    \epsilon \cos{\chiup}}} {2 \pi I_{0}
    (\epsilon)},
    \label{P_eq}
\end{equation}
where the dimensionless parameter
\begin{equation}
    \epsilon = \frac{\alpha}{\beta^{2}} =
    \frac{M H_{m}V}{k_{B}T}
    \label{def_eps}
\end{equation}
is the ratio of magnetic to thermal energy and $I_{0}(\epsilon)$ is the modified Bessel function of the first kind and order zero. Note, since the von Mises distribution is the circular analog of the normal distribution, it is widely used in various applications, see, e.g., Refs.~\cite{Mard1972, JaSe2001}.

As it follows from (\ref{<s_y>1}), in the planar rotor model the drift velocity of ferromagnetic particles is given by
\begin{equation}
    \langle s_{y} \rangle = \gamma
    \int_{0}^{1} \langle \sin{\chi_{
    \mathrm{st}}(\xi)}\rangle \sin{
    (2\pi \xi - \phi)}\, d\xi
    \label{<s_y>2}
\end{equation}
and the mean value of $\sin{\chi_{\mathrm{st}}(\xi)}$ is defined as
\begin{equation}
    \langle \sin{\chi_{\mathrm{st}}(\xi)}
    \rangle = \int_{0}^ {2\pi}\sin{\chiup}
    P_{\mathrm{st}}(\chiup, \xi)\, d\chiup
    \label{<sin>1}
\end{equation}
with $P_{\mathrm{st}}(\chiup, \xi) = \lim_{n \to \infty} P(\chiup, n + \xi)$. Our aim here is to determine the drift velocity (\ref{<s_y>2}) in the limit of large thermal fluctuations, when $\beta \to \infty$.

Since, according to (\ref{def_eps}), in this limit $\epsilon \to 0$, the steady-state solution of Eq.~(\ref{FP3}) can be represented in the form
\begin{equation}
    P_{\mathrm{st}}(\chiup, \xi) = \sum
    _{l=0}^{\infty} P_{l}(\chiup, \xi),
    \label{rep1}
\end{equation}
where the functions $P_{l}(\chiup, \xi) = P_{l}$ are of the order of $\epsilon^{l}$, $P_{l} \sim \epsilon^{l}$, and satisfy the periodicity conditions $P_{l}(\chiup + 2\pi, \xi) = P_{l}(\chiup, \xi)$ and $P_{l}(\chiup, \xi + 1) = P_{l}(\chiup, \xi)$. In the zeroth-order approximation one obtains $P_{0} = \lim_{\epsilon \to 0} P_{\mathrm{eq}}(\chiup) = 1/2\pi$, and the functions $P_{l}$ ($l \geq 1$) are connected with the functions $P_{l-1}$ as follows:
\begin{equation}
    \frac{\partial^{2}} {\partial
    \chiup^{2}} P_{l} = \frac{\epsilon}
    {\alpha} \left(\frac{\partial}
    {\partial \xi} P_{l-1}
    + \dot{\psi}(\xi) \frac{\partial}
    {\partial \chiup} P_{l-1} - \alpha
    \frac{\partial}{\partial \chiup}
    \sin{\chiup} P_{l-1} \right).
    \label{eq_P_l}
\end{equation}

Using this recurrence equation, it is not difficult to show that the steady-state probability density (\ref{rep1}) in the second-order approximation (when $l \leq 2$) is given by
\begin{equation}
    P_{\mathrm{st}}(\chiup, \xi) = \frac{1}
    {2\pi} + \frac{\epsilon}{2\pi} \cos{\chiup}
    + \frac{\epsilon^{2}}{8\pi} \cos{2\chiup}
    + \frac{\epsilon^{2}}{2\pi \alpha} \dot{
    \psi}(\xi) \sin{\chiup}.
    \label{P_st1}
\end{equation}
The first three terms on the right-hand side of this expression represent the second-order expansion of $P_{\mathrm{eq}}(\chiup)$ and, according to (\ref{<s_y>2}) and (\ref{<sin>1}), they do not contribute to the drift velocity $\langle s_{y} \rangle$. In contrast, the last term in (\ref{P_st1}), which accounts for the particle rotation induced by the magnetic field, yields
\begin{equation}
    \langle s_{y} \rangle = \frac{\gamma
    \epsilon^{2}}{2 \alpha} \int_{0}^{1}
    \dot{\psi}(\xi)\sin{(2\pi\xi - \phi)}
    \,d\xi.
    \label{<s_y>3}
\end{equation}

Let us assume for simplicity that the azimuthal angle $\psi(\tau)$ of the magnetic field (\ref{def_h}) changes with time as
\begin{equation}
    \psi(\tau) =\psi_{m} \cos{(2\pi\tau)},
    \label{psi}
\end{equation}
where the positive parameter $\psi_{m}$ is measured in radians and shows how many rotations the magnetic field does in clockwise and counterclockwise directions during one period of the field. Then, taking into account that in this case
\begin{equation}
    \int_{0}^{1} \dot{\psi}(\xi)\sin{
    (2\pi \xi - \phi)}\, d\xi = - \pi
    \psi_{m} \cos{\phi},
    \label{int}
\end{equation}
the particle drift velocity (\ref{<s_y>3}) reduces to
\begin{equation}
    \langle s_{y} \rangle = -\frac{\pi
    \psi_{m}\gamma\epsilon^{2}}{2 \alpha}
    \cos{\phi}.
    \label{<s_y>4}
\end{equation}

According to this result, strong thermal fluctuations essentially decrease the drift velocity ($\langle s_{y} \rangle$ tends to zero quadratically as $\epsilon \to 0$). In addition, it shows that $\langle s_{y} \rangle$ depends linearly on the amplitude $\psi_{m}$ of the magnetic field azimuthal angle and periodically on the initial phase $\phi$ of the external force.

\subsection{General model}
\label{Gen}

Because the deterministic dynamics of the particle magnetization occurs in the plane of magnetic field rotation \cite{DePe2017, DPKD2017}, the planar rotor model is quite appropriate to describe the drift motion of ferromagnetic particles in the absence of thermal fluctuations. However, the presence of such fluctuations, especially strong ones, casts doubt on the applicability of this model. The reason is that the out-of-plane fluctuations of the magnetization vector, which always exist, are excluded from consideration in the planar rotor model. Therefore, to investigate the influence of these fluctuations on the drift velocity, we will use a general Fokker-Planck equation (\ref{FP}).

In the case of static magnetic field (when $\psi(\tau) = \mathrm{ const}$) the equilibrium solution of Eq.~(\ref{FP}) has a well-known form
\begin{equation}
    P_{\mathrm{eq}}(\thetaup, \chiup) =
    \frac{1} {4\pi} \frac{\epsilon}
    {\sinh{\epsilon}}\sin{\thetaup}\,
    e^{\epsilon \sin{\thetaup}
    \cos{\chiup}}.
    \label{P_eq2}
\end{equation}
At the same time, if $\psi(\tau) \neq 0$ and $\epsilon \to 0$, the steady-state solution of this equation can be represented as
\begin{equation}
    P_{\mathrm{st}}(\thetaup, \chiup, \xi) =
    \sum_{l=0}^{\infty} P_{l}(\thetaup,
    \chiup, \xi),
    \label{rep2}
\end{equation}
where $P_{l}(\thetaup, \chiup, \xi) = P_{l} \sim \epsilon^{l}$ and $\int_{0}^{\pi} \int_{0}^{2\pi} P_{l}\, d\thetaup d\chiup = 0$ for $l \geq 1$. From (\ref{P_eq2}) one immediately finds $P_{0} = \lim_{\epsilon \to 0} P_{\mathrm{eq}}(\thetaup, \chiup) = (1/4\pi) \sin{\thetaup}$, and using Eq.~(\ref{FP}), it is not difficult to show that, if term $P_{l-1}$ (with $l \geq 1$) is known, then $P_{l}$ can be determined by solving the following equation:
\begin{align}
    \frac{\partial^{2}}{\partial
    \thetaup^{2}} P_{l}
    & + \frac{1} {\sin^{2}{\thetaup}}
    \frac{\partial^{2}}{\partial \chiup^{2}}P_{l} - \frac{\partial}{\partial
    \thetaup} \cot{\thetaup} P_{l} = \frac{\epsilon}{\alpha}
    \left(\frac{\partial}{\partial\tau} P_{l-1} + \dot{\psi} \frac{\partial}{
    \partial\chiup} P_{l-1}\right.
    \nonumber \\
    & + \left. \alpha \cos{\chiup}
    \frac{\partial} {\partial \thetaup}
    \cos{\thetaup} P_{l-1} - \alpha \frac{1}{
    \sin{\thetaup}}\frac{\partial}{
    \partial \chiup}\sin{\chiup} P_{l-1}\right).
    \label{FP4}
\end{align}

The sequential solution of this equation for $l =1$ and 2 permitted us to determine the steady-state probability density $P_{\mathrm{st}} (\thetaup, \chiup, \xi)$ with the second-order accuracy
\begin{align}
    P_{\mathrm{st}}& (\thetaup, \chiup, \xi)
    = \frac{1}{4\pi} \sin{\thetaup} +
    \frac{\epsilon}{4\pi} \sin^{2}{\thetaup}
    \cos{\chiup} - \frac{\epsilon^{2}}{24\pi}
    \sin{\thetaup}
    \nonumber \\
    & \times (1- 3 \sin^{2}{\thetaup} \cos^{2}{\chiup})
    + \frac{\epsilon^{2}}{8\pi \alpha} \dot{
    \psi}(\xi) \sin^{2}{\thetaup} \sin{\chiup}.
    \label{P_st2}
\end{align}
Like in the previous case, the first three terms on the right-hand side of this formula arise from the expansion of the equilibrium probability density (\ref{P_eq2}) and, as it follows from the definition
\begin{equation}
    \langle \sin{\theta_{\mathrm{st}}(\xi)} \sin{\chi_{\mathrm{st}}(\xi)}\rangle =
    \int_{0}^{\pi} \int_{0}^{2\pi} \sin{
    \thetaup} \sin{\chiup}P_{\mathrm{st}}
    (\thetaup, \chiup, \xi)\, d\thetaup
    d\chiup,
    \label{<sin-sin>1}
\end{equation}
they do not contribute to the drift velocity $\langle s_{y} \rangle$. In contrast, since $\int_{0}^{\pi} \sin^{3}{\thetaup}\, d\thetaup = 4/3$ and $\int_{0}^{2\pi} \sin^{2}{\chiup}\, d\chiup = \pi$, the last term in (\ref{P_st2}) does contribute to $\langle s_{y} \rangle$, yielding
\begin{equation}
    \langle s_{y} \rangle = \frac{\gamma
    \epsilon^{2}}{6 \alpha} \int_{0}^{1}
    \dot{\psi}(\xi)\sin{(2\pi\xi - \phi)}
    \, d\xi
    \label{<s_y>5}
\end{equation}
for arbitrary $\psi(\tau)$ and
\begin{equation}
    \langle s_{y} \rangle = -\frac{\pi
    \psi_{m}\gamma\epsilon^{2}}{6 \alpha}
    \cos{\phi},
    \label{<s_y>6}
\end{equation}
if $\psi(\tau)$ is defined by (\ref{psi}).

By comparing formulas (\ref{<s_y>3}) and (\ref{<s_y>5}), we see that the drift velocity predicted by the planar rotor model in the case of large thermal fluctuations is three times larger than that predicted by the general model. Because the only difference between these models is the out-of-plane fluctuations of the magnetization vector, we may conclude that these fluctuations are responsible for this decrease of the drift velocity.

\section{Temperature effects in the drift motion: Numerical results}
\label{Num}

In order to verify our analytical results (\ref{<s_y>4}) and (\ref{<s_y>6}) for the drift velocity, we first performed numerical simulations of the stochastic equations (\ref{eq_chi}) and (\ref{red_set}) and determined their solutions, $\chi^{(i)} (n + \xi)$ and $\theta^{(i)} (n + \xi)$ and $\chi^{(i)} (n + \xi)$, in the $i$th run. Then, choosing $n \gg 1$, we  approximately calculated the means $\langle \sin{\chi_{\mathrm{ st}}(\xi)} \rangle$ and $\langle \sin{\theta_{ \mathrm{st}}(\xi)} \sin{\chi_{ \mathrm{st}} (\xi)}\rangle$ as
\begin{equation}
    \langle \sin{\chi_{\mathrm{st}}(\xi)}
    \rangle = \frac{1}{N} \sum_{i=1}^{N}
    \sin{\chi^{(i)} (n + \xi)}
    \label{def_<sin>}
\end{equation}
and
\begin{equation}
    \langle \sin{\theta_{\mathrm{st}}(\xi)}
    \sin{\chi_{\mathrm{st}}(\xi)}\rangle =
    \frac{1}{N} \sum_{i=1}^{N}\sin{\theta^
    {(i)} (n + \xi)} \sin{\chi^{(i)} (n +
    \xi)},
    \label{def_<sinsin>}
\end{equation}
where $N(\gg 1)$ is the total number of runs (these formulas become exact if $N, n \to \infty$). Finally, using (\ref{def_<sin>}) and (\ref{def_<sinsin>}), we determined the drift velocity (\ref{<s_y>2}) within the planar rotor model and the drift velocity (\ref{<s_y>1}) within the general model.

\begin{figure}[ht]
    \centering
    \includegraphics[width=\columnwidth]{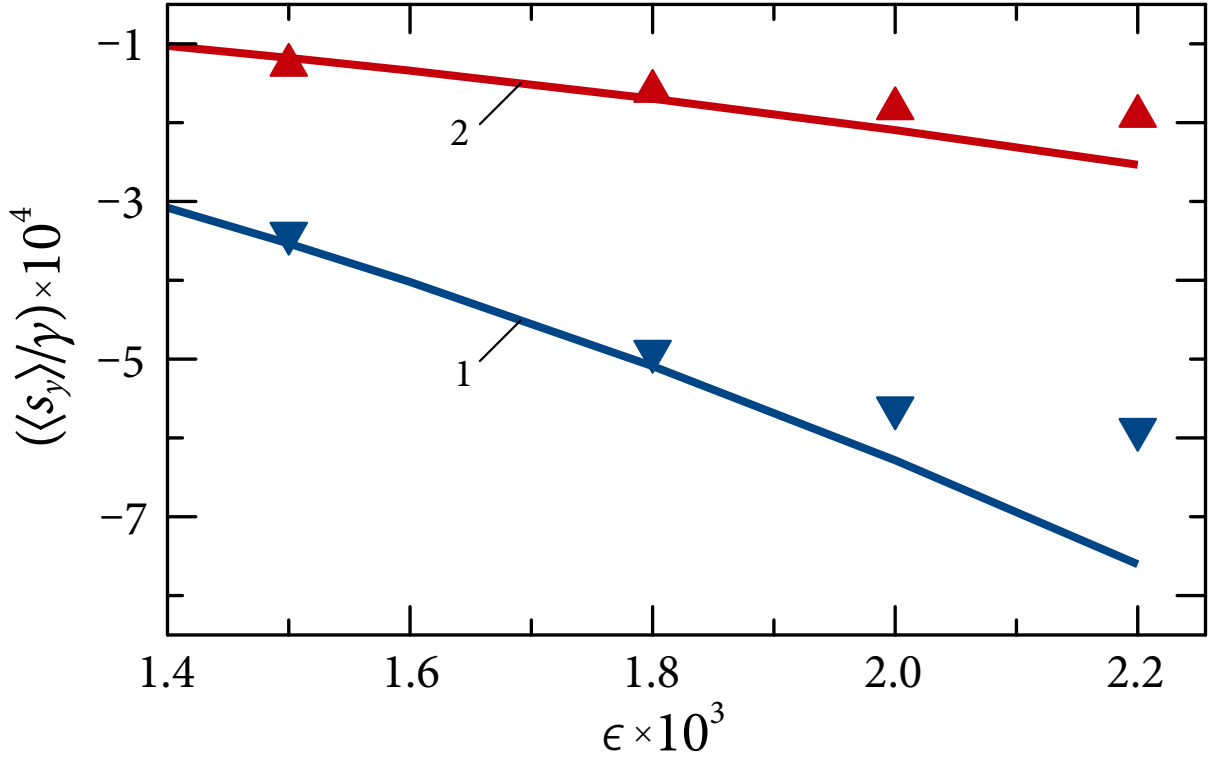}
    \caption{The reduced drift velocity as a
    function of the parameter $\epsilon$ for
    $\epsilon \ll 1$. The inverted and upright
    triangles represent the simulation data
    for the planar rotor and general models,
    and the solid lines 1 and 2 represent the
    theoretical results from (\ref{<s_y>4})
    and (\ref{<s_y>6}), respectively.}
    \label{fig1}
\end{figure}

The reduced drift velocity $\langle s_{y} \rangle / \gamma$ as a function of $\epsilon$, obtained by the above procedure for two models, is shown in Fig.~\ref{fig1} (the simulation parameters are chosen to be $\alpha = 0.01$, $\psi_{m} = 1\, \mathrm{rad}$, $\phi = 0\, \mathrm{rad}$, $n= 200$, and $N= 10^{7}$). As seen, the simulation data are in good
agreement with analytical predictions, if the parameter $\epsilon$ is small enough. (The difference between the numerical and analytical results grows with increasing $\epsilon$ because of limitation of the second-order approximation.) This validates both our analytical results and simulation procedure. Moreover, the obtained numerical results confirm a strong influence of the out-of-plane fluctuations of the particle magnetization on the drift velocity. Therefore, to account for these fluctuations, in our further analysis we use a general set of effective stochastic differential equations (\ref{red_set}).

It should be noted that, according to Eqs.~(\ref{red_set}), the influence of temperature on the rotational properties of particles may occur not only because of the parameter $\beta$ explicitly depends on $T$, see (\ref{def_a,b}), but also due to the temperature dependence of the magnetization $M$ and dynamic viscosity $\eta$. For illustration, we consider $\mathrm{Co}$ nanoparticles suspended in water whose temperature varies in the interval $\Delta T$ ranging from $273\, \mathrm{K}$ (near-freezing temperature) to $373\, \mathrm{K}$ (near-boiling temperature). Since the Curie temperature of cobalt is $1.398 \times 10^{3}\, \mathrm{K}$ (see Ref.~\cite{Gui2017}, Tab.~2.3), the $\mathrm{Co}$ magnetization slowly varies in the interval $\Delta T$. Therefore, its value in this interval can be chosen to be constant, e.g., $M= 1.422 \times 10^{3}\, \mathrm{emu /cm^{3}}$, which is the $\mathrm{Co}$ saturation magnetization at room temperature (see Ref.~\cite{Gui2017}, Tab.~2.9). In contrast, the dynamic viscosity of water strongly decreases with increasing temperature. For example, in accordance with the experimental data (see Ref.~\cite{VGPD2007}, Chap.~6), one obtains $\eta|_{T = 273.64\, \mathrm{K}}/ \eta|_{T = 371.24\, \mathrm{K}} = 6.095$. As the parameters $\alpha$ and $\beta$ depend on $\eta$, this means that the temperature dependence of the dynamic viscosity must be taken into account to correctly describe the drift motion of these nanoparticles. For this purpose, we use the approximate formula
\begin{equation}
    \eta = 10^{A + B/(T-C)},
    \label{appr_eta}
\end{equation}
which well reproduces the experimental data for the dynamic viscosity of water in the reference interval $\Delta T$ (see Ref.~\cite{VGPD2007}, Tab.~4.9). Here, $A = - 3.5318$, $B = 220.57$, $C = 149.39$, $T$ is in Kelvin, and $\eta$ is in poise (P).

\begin{figure}[ht]
    \centering
    \includegraphics[width=\columnwidth]{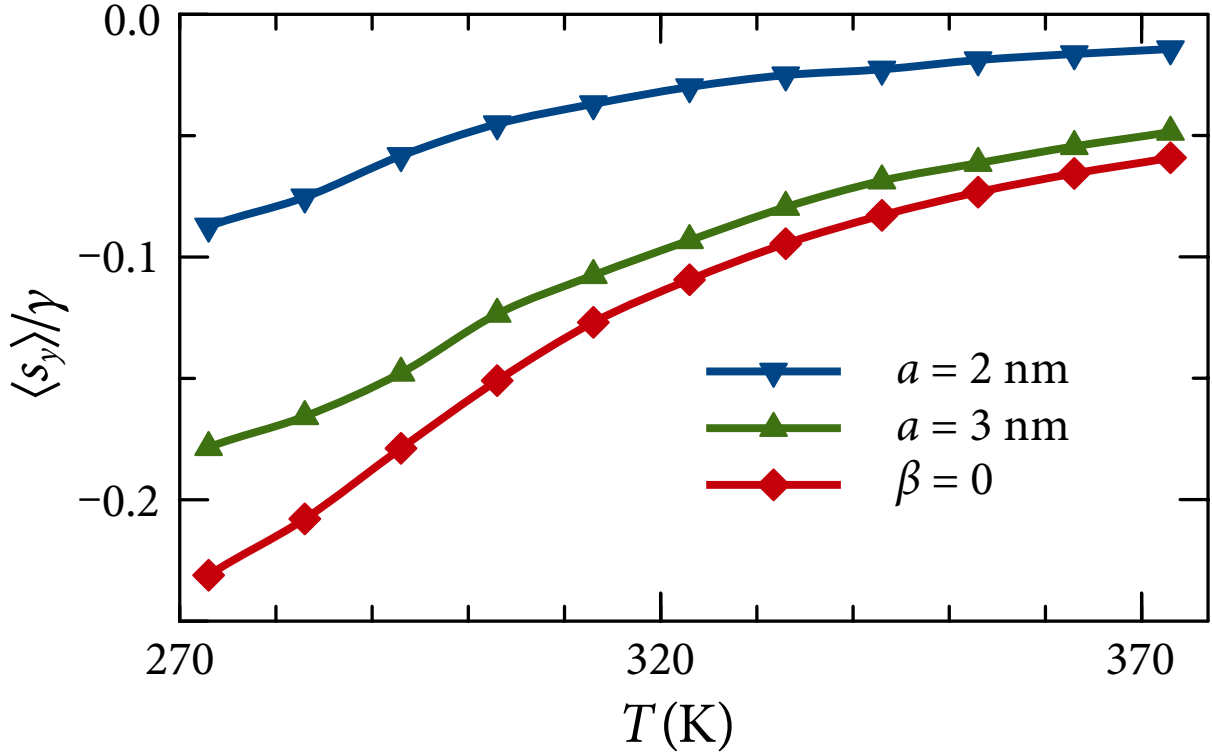}
    \caption{The temperature dependence of the
    reduced drift velocity for $\mathrm{Co}$
    nanoparticles of different radii suspended
    in water. The dependence at $\beta =0$
    corresponds to the large radius limit.}
    \label{fig2}
\end{figure}

Using the above described procedure, we numerically determined the reduced drift velocity $\langle s_{y} \rangle / \gamma$ as a function of temperature for $\mathrm{Co}$ nanoparticles suspended in water (in all simulations we take $n=50$ and $N=10^{5}$). Figure \ref{fig2} illustrates this function in the cases when $H_{m} = 10^{2}\, \mathrm{Oe}$, $\psi_{m} = 1\, \mathrm{rad}$, $\Omega = 10^{6}\, \mathrm{rad/s}$, $\phi = 0\, \mathrm{rad}$,
$a = 2\, \mathrm{nm}$ (line with inverted triangles), and $a = 3\, \mathrm{nm}$ (line with upright triangles). With increasing the particle radius $a$, the temperature dependence of the drift velocity very rapidly approaches the limiting function (represented by the line with diamonds), which corresponds to the noiseless case $\beta =0$. In particular, the difference between these functions can safely be neglected already at $a > 5\, \mathrm{nm}$ (this radius slowly increases with decreasing $H_{m}$). This means that thermal fluctuations reduce the drift velocity of rather small particles (when $a < 5\, \mathrm{nm}$), while in the case of larger particles their drift velocity depends on temperature solely due to the temperature dependence of the dynamic viscosity of water.

According to (\ref{<s_y>1}), the drift velocity $\langle s_{y} \rangle$ is a periodic function of the initial phase $\phi$ satisfying the condition $\langle s_{y} \rangle |_{\pi + \phi} = - \langle s_{y} \rangle |_{\phi}$. As a consequence, $\langle s_{y} \rangle$ at fixed temperature changes sign (i.e., the drift direction is reversed) by changing $\phi$. Since the drift velocity of particles of any size depends on temperature (see above), one
\begin{figure}[ht]
    \centering
    \includegraphics[width=\columnwidth]{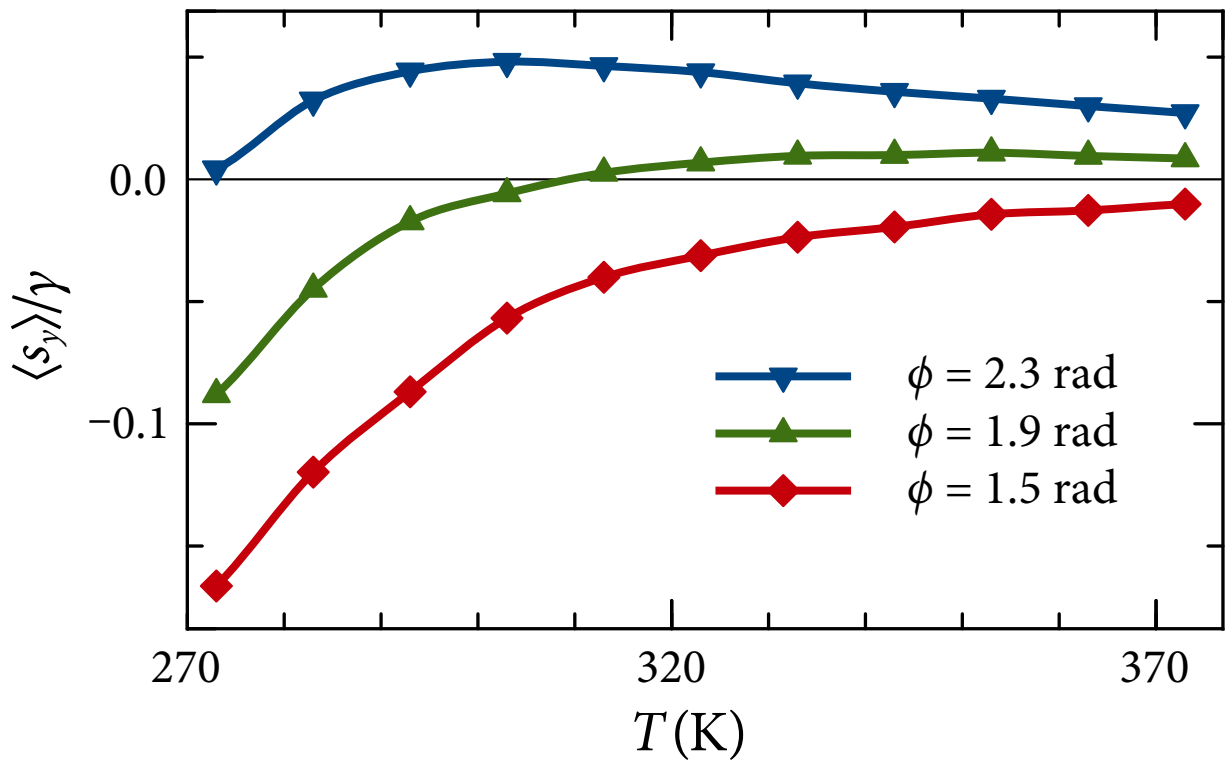}
    \caption{The reduced drift velocity of
    $\mathrm{Co}$ nanoparticle as a function
    of temperature for different values of the
    initial phase $\phi$. The particle radius is
    $a = 3\, \mathrm{nm}$ and the other
    parameters are the same as in Fig.~\ref{fig2}.}
    \label{fig3}
\end{figure}
may expect that temperature change can also change the sign of $\langle s_{y} \rangle$, if the initial phase is properly chosen. This expectation is confirmed by our simulation results presented in Fig.~\ref{fig3}. From a physical point of view, the reversal of the drift direction occurs when the Magnus force averaged over the time period and thermal fluctuations changes sign, which in turn happens if the translational and rotational oscillations of particles are appropriately synchronized. This synchronization can be realized by changing both the initial phase (that is quite obvious) and temperature. It seems that the observed temperature-induced reversal of the drift direction can be used to improve the method of separation of suspended ferromagnetic particles based on the Magnus effect \cite{DePe2017}.

\begin{figure}[ht]
    \centering
    \includegraphics[width=\columnwidth]{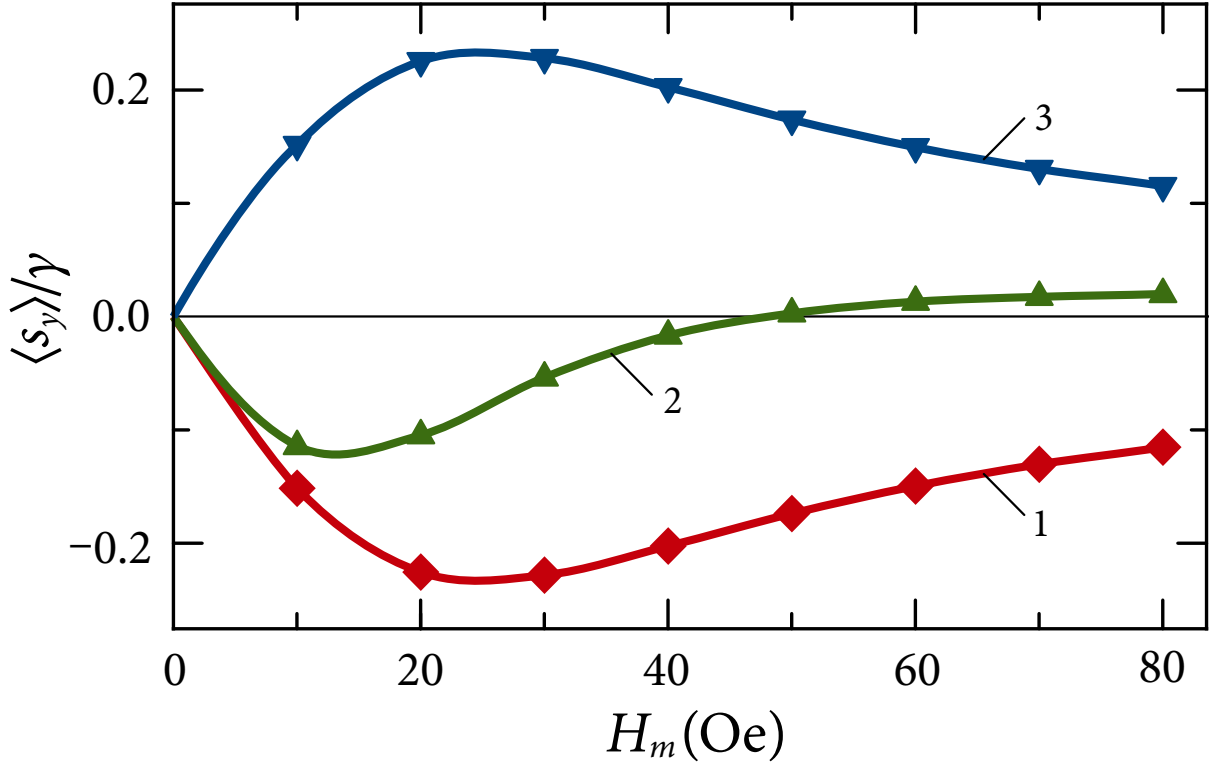}
    \caption{The reduced drift velocity of
    $\mathrm{Co}$ nanoparticle as a function
    of the magnetic field magnitude $H_{m}$
    for different values of the initial
    phase $\phi$. Line 1 corresponds to
    $\phi = 0.4\, \mathrm{rad}$, line 2 to
    $\phi = 2.0\, \mathrm{rad}$, and line
    3 to $\phi = (\pi + 0.4)\, \mathrm{rad}$.}
    \label{fig4}
\end{figure}

\begin{figure}[ht]
    \centering
    \includegraphics[width=\columnwidth]{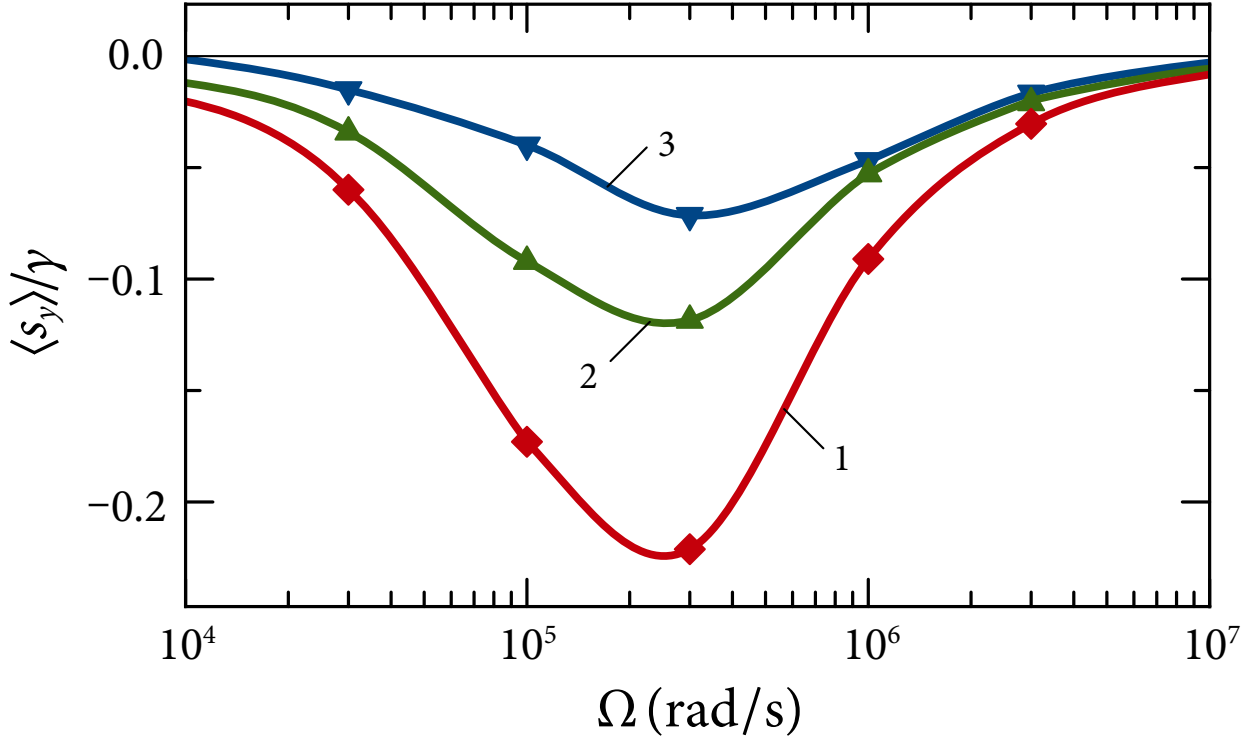}
    \caption{The frequency dependence of the
    reduced drift velocity for different radii
    of $\mathrm{Co}$ nanoparticles. Line 1
    corresponds to the large radius limit
    ($\beta = 0$), line 2 to $a = 5\,
    \mathrm{nm}$, and line 3 to $a = 4\,
    \mathrm{nm}$. The $\Omega$-axis is
    in logarithmic scale.}
    \label{fig5}
\end{figure}

\begin{figure}[ht]
    \centering
    \includegraphics[width=\columnwidth]{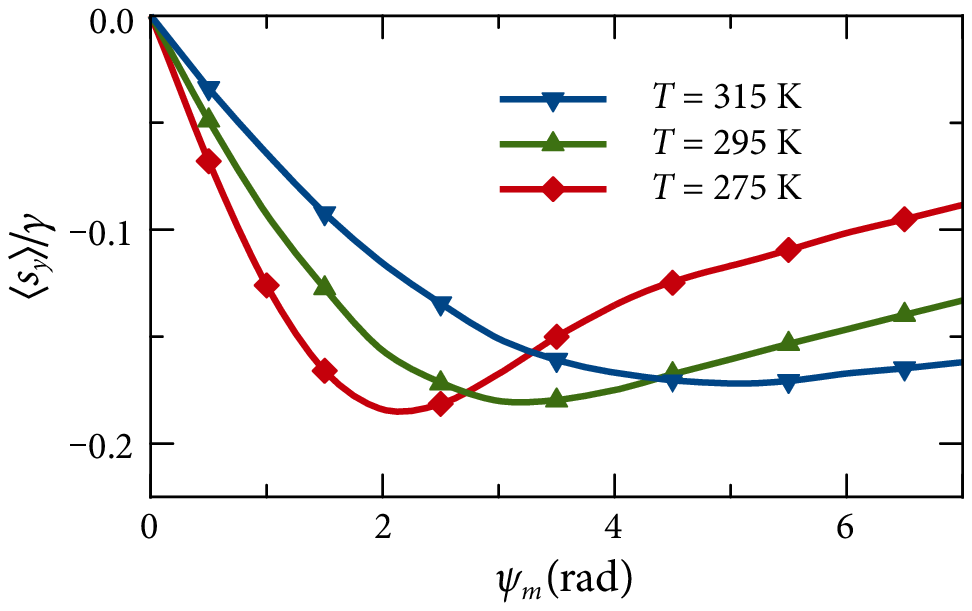}
    \caption{The reduced drift velocity of
    $\mathrm{Co}$ nanoparticle as a function
    of the angle $\psi_{m}$ for different
    temperatures.}
    \label{fig6}
\end{figure}

In order to gain more insight into the behaviour of the reduced drift velocity of $\mathrm{Co}$ nanoparticles, we calculated also its dependence on the magnetic field characteristics, namely, on the magnitude $H_{m}$, frequency $\Omega$, and angle $\psi_{m}$. The dependence of $\langle s_{y} \rangle / \gamma$ on $H_{m}$ for a nanoparticle of radius $a = 5\, \mathrm{nm}$ determined at room temperature $T = 295\, \mathrm{K}$, $\Omega = 5\times 10^{5}\, \mathrm{rad/s}$, $\psi_{m} = 1\, \mathrm{rad}$, and different values of $\phi$ is shown in Fig.~\ref{fig4}. Since the change of $H_{m}$ is equivalent to changing the parameter $\alpha$, the obtained results are similar to those derived in Ref.~\cite{DPKD2017} in the deterministic approach. Figure \ref{fig5} illustrates the dependence of $\langle s_{y} \rangle / \gamma$ on $\Omega$ calculated for $T = 295\, \mathrm{K}$, $H_{m} = 10\, \mathrm{Oe}$, $\psi_{m} = 1\, \mathrm{rad}$, $\phi = 0\, \mathrm{rad}$, and different particle radii. As seen, decreasing the particle size decreases the absolute value of the reduced drift velocity and shifts the minimum of this velocity to the right. This occurs because the magnitude of thermal fluctuations, which is characterized by the parameter $\beta$, increases as the particle size decreases, see (\ref{def_a,b}). Finally, in Fig.~\ref{fig6} we show the dependence of $\langle s_{y} \rangle / \gamma$ on $\psi_{m}$ calculated for $a = 5\, \mathrm{nm}$, $H_{m} = 10\, \mathrm{Oe}$, $\Omega = 5\times 10^{5}\, \mathrm{rad/s}$, $\phi = 0\, \mathrm{rad}$, and different temperatures. In this case, the influence of temperature on the reduced drift velocity is caused by both the temperature dependence of the dynamic viscosity of water and thermal fluctuations. It should also be kept in mind that the dependencies of the dimensional drift velocity $v_{\mathrm{dr}} = v_{m}\langle s_{y} \rangle$ on temperature differ from that obtained for the reduced drift velocity $\langle s_{y} \rangle / \gamma$. The reason is that the quantities $v_{m}$ and $\gamma$, according to their definitions, depend on $T$ through the dynamic viscosity of water.

\section{Conclusions}
\label{Concl}

We have studied analytically and numerically the temperature dependence of the drift velocity of suspended single-domain ferromagnetic particles induced by the Magnus force. Our approach is based on a set of stochastic differential equations that describes the translational and rotational motions of particles with frozen magnetization in the limit of small Reynolds numbers. In this approximation, the rotational motion, which is caused by a rotating magnetic field, affects (due to the Magnus effect) the translational motion, which is caused by an external harmonic force. In contrast, the translational motion does not influence the rotational one. These features of particle dynamics were used to derive a general expression for the particle drift velocity in the steady state. It shows that, as in the deterministic case \cite{DePe2017, DPKD2017}, drift occurs when the translational and rotational motions of particles are properly synchronized, and thermal fluctuations do not destroy the drift motion; they only decrease the drift velocity. Thus, the statistical theory developed here confirms the existence of the drift phenomenon in suspensions of ferromagnetic particles and has a much wider range of applicability compared to the deterministic one.

Using the planar rotor model and general model, we have solved the corresponding Fokker-Planck equations in the limit of strong thermal fluctuations and, on this basis, calculated the particle drift velocity for these models. It turned out that the out-of-plane fluctuations, which are not accounted for by the planar rotor model, play an important role, resulting in decreasing the drift velocity by three times.

To verify our analytical results obtained in the case of large thermal fluctuation and to analyze the dependence of the drift velocity on temperature and other parameters, we have solved numerically a set of effective stochastic differential equations describing the magnetization dynamics. In this way, it has been shown that the analytical and numerical results are in good agreement with each other, if the magnetic energy of a particle is very small compared to the thermal energy. Next, using the validated numerical procedure, we have calculated the temperature dependence of the drift velocity of $\mathrm{Co}$ nanoparticles suspended in water. An important feature of our approach is that the temperature dependence of the dynamic viscosity of water is explicitly taken into account. This permitted us to determine the drift velocity in a wide temperature interval and clarify the role of thermal fluctuations. It has been shown, in particular, that the temperature dependence of the drift velocity of rather large particles is almost completely determined by the temperature dependence of the dynamic viscosity of water, and thus thermal fluctuations do not noticeably affect the drift velocity of these particles. In contrast, thermal fluctuations strongly influence (decrease) the drift velocity of smaller particles whose size is of the order of a few nanometers or less. But the most remarkable feature of the drift velocity is that its sign (i.e., the drift direction) can be changed by changing temperature, if the initial phase of the external driving force is chosen appropriately. This opens new perspectives in the development of innovative methods for separation of ferromagnetic particles in suspensions.

\section*{Acknowledgments}

This work was supported by the Ministry of Education and Science of Ukraine under Grant No.\ 0116U002622.

\section*{References}

\bibliographystyle{plain}

\end{document}